# Violation of Bell inequality by four-photon Greenberger–Horne–Zeilinger state with a phase from a warm atomic ensemble

Jiho Park[1], Junghee Ryu[2, †], Heonoh Kim[1], and Han Seb Moon[1, *]

[1] *Department of Physics, Pusan National University, Geumjeong-Gu, Busan 46241, Republic of Korea*
[2] *Division of National Supercomputing, Korea Institute of Science and Technology Information, Daejeon 34141, Republic of Korea*
[†] *rjhui82@gmail.com*
[*] *hsmoon@pusan.ac.kr*

A Greenberger–Horne–Zeilinger (GHZ) entangled state with a phase is crucial for realizing desired multipartite quantum states for practical applications. Here, we report violations of the general Bell inequality (GBI) introduced in [1] using the four-photon polarization-entangled phase-GHZ state realized via intrinsic polarization correlation and collective two-photon coherence in the $5S_{1/2}$–$5P_{3/2}$–$5D_{5/2}$ transition of $^{87}$Rb atoms. The phase-GHZ state can be achieved by the unitary transformation of only one local phase of the four photons. Theoretically, the GHZ state with the $\pi/4$ phase affords maximal violation of the GBI of $2\sqrt{2}$ at the local measurement settings of the Pauli operators $\sigma_x$ and $\sigma_y$. We experimentally demonstrate strong violations of the GBI of the phase-GHZ state by 47 standard deviations. In addition to the entanglement witness for the phase-GHZ state, the results represent a genuine four-photon entanglement of the phase-GHZ state, thereby providing a novel resource for realizing photonic quantum computation, magic-state distillation from entangled states, and quantum networks based on atom–photon interactions.

## INTRODUCTION

Multipartite quantum states are of fundamental importance for gaining greater understanding of the relationship between classical and quantum physics [2]. These states are essential for fundamental research on quantum physics and quantum information processing applications such as quantum computing, quantum metrology, and quantum communication [3-9]. In particular, Greenberger–Horne–Zeilinger (GHZ) states are the best-understood multipartite quantum states that exhibit the behavior of perfect quantum correlations in a different way from other bipartite states [10]. For example, the realization of the GHZ state allows one to directly refute the ideas of Einstein–Podolsky–Rosen on the relation between quantum mechanics and the local realistic description [11], which is well known as "all versus nothing" test based on perfect correlations of the GHZ states. The generation of multiphoton GHZ states has been experimentally demonstrated in various physical systems such as nonlinear crystals, silicon nanowires, and atomic ensembles [12-17]. To realize photonic quantum information processes and long-distance quantum networks with quantum memories, quantum nodes, and quantum repeaters based on atom–photon interactions, the multiphoton GHZ entangled state generated from atomic ensembles is a crucial quantum resource [16-18].

In this context, a narrowband polarization-entangled four-photon GHZ state with a spectral bandwidth of 19.5 MHz, generated from a cold atomic ensemble, has been reported using two polarization-entangled photon pairs through a double-lambda configuration [16]. However, the brightness, robustness, and fidelity of the entangled multiphoton source generated from the cold atomic ensemble are currently unsatisfactory. Recently, to overcome the abovementioned problems, researchers have experimentally demonstrated a four-photon polarization-entangled GHZ state with a high generation rate and high fidelity via the spontaneous four-wave mixing (SFWM) process occurring in a warm atomic ensemble [17-18]. Furthermore, the stable generation of a bright four-photon polarization-entangled GHZ state has been successfully achieved by spatial multiplexing of inherently polarization-entangled photons from a warm atomic ensemble via the polarization correlation of the decay channels of the Zeeman substates in a cascade-type atomic system [18].

The two-photon polarization-entangled Bell states from atomic ensembles have been shown to strongly violate the Clauser–Horne–Shimony–Holt (CHSH) form of the Bell inequality, with the *S*-value (the quantum violation) reaching 2.65 by 65 standard deviations [19-20]; this result is a confirmation of the polarization entanglement of the resource generated from the atomic ensemble. In general, it is well known that Bell inequalities can be deployed as an entanglement certification; that is, quantum violations are necessary and sufficient conditions for the detection of entanglement in a device-independent manner [21].

Multipartite systems can have specific traits such as genuine multipartite entanglement, which are not present in bipartite cases. As the Bell inequalities for multipartite systems exhibit significantly more abundant and complex structures than those of bipartite systems, the behaviors of quantum violations are nontrivial [21-23]. In addition, it is difficult to determine whether genuine multipartite entanglement has been realized experimentally. In this context, the Mermin–Ardehali–Belinskii–Klyshko (MABK) inequalities for $N$ particles were proposed [24-26] and experimentally demonstrated with multiphoton GHZ states via the spontaneous parametric down-conversion (SPDC) process in a nonlinear crystal [14]. Furthermore, Zukowski and Brukner derived a single general Bell inequality (GBI) for $N$ particles and showed quantum violations using the $N$-partite GHZ state [1]. Despite the successful realization of a multiphoton GHZ entangled state generated from an atomic ensemble, violations of the Bell-type inequality of a multiphoton GHZ state from a warm atomic ensemble have not been reported thus far. Thus, in this study, we demonstrate experimental violations of the Bell inequality using the entangled state generated from an atomic ensemble, which can guarantee the quantum entanglement of our resource. In addition, we examine whether our resource affords genuine multipartite entanglement by measuring a suitable entanglement witness.

**RESULTS**

**Four-photon GHZ state with a phase from a warm atomic ensemble**

We demonstrate the violation of the GBI using a four-photon polarization-entangled GHZ state with a highly stable local phase, which is generated via the SFWM process occurring in a warm atomic ensemble of $^{87}$Rb. In our experimental system, the phase of the four-photon polarization-entangled GHZ state can be stably controlled by the unitary transformation of the local phase by tuning the tilt angle of the quarter-wave plate positioned along the path of one of the four photons. Owing to the stable generation of a bright four-photon polarization-entangled GHZ state, we find that the GHZ state with a phase of $\pi/4$ gives rise to a maximal violation of the GBI in the linear- and circular-polarization measurements. The multipartite GHZ state with phase has also been used to show quantum violations of the Mermin–Klyshko inequality [27]. Recently, it has been theoretically shown that entangled states with a phase can be deployed for magic-state distillation, which in general the entanglement is regarded as a disadvantage for the distillation protocol [28]. In particular, multipartite GHZ-type states with phases can improve the success probability of the magic-state distillation protocol. We also discuss the violation of GBI as an entanglement criterion for four-photon GHZ states according to the local phase. Furthermore, the four-photon GHZ state with a local phase is fully characterized using quantum-state tomography (QST).

Figure 1 shows the configuration for entangled photon-pair generation from the $5S_{1/2}$–$5P_{3/2}$–$5D_{5/2}$ transition of $^{87}$Rb atoms and the experimental setup for polarization-entangled four-photon GHZ state, realized by combining photons from two different spatial modes of the polarization-entangled photon-pair source. Owing to angular momentum conservation in the atom–photon interaction, the polarizations of the signal and idler photons are strongly correlated with the $|H\rangle_s|V\rangle_i$ and $|V\rangle_s|H\rangle_i$ states, respectively, where the transition probabilities of both decay channels in the Zeeman substates are equal upon considering the $\pi$ and $\sigma^{\pm}$ transitions. Therefore, the entangled two-photon state in our system can be expressed as a maximally entangled Bell state as follows:

$$|\Psi^+\rangle = \frac{1}{\sqrt{2}}\left(|H\rangle_s|V\rangle_i + |V\rangle_s|H\rangle_i\right). \quad (1)$$

A four-photon GHZ state can be generated by combining the two idler photons from the different symmetrical spatial modes of the polarization-entangled photon pairs, where the two modes are denoted as Pair$_1$ (yellow) and Pair$_2$ (green), as shown in Fig. 1(b). The numbers along each path can be understood as follows: 1 and 4 denote the input paths of the two signal photons, and 2 and 3 denote the polarizing beam splitter (PBS) output paths of the idler photons. Owing to the uniqueness of the atom, the biphoton waveforms of both two-photon states in Fig. 1(b) are spectrally and temporally identified; however, the characteristics of entangled photons from the atomic ensembles are identified and intrinsically indistinguishable. Furthermore, in our experiment, because we did not use any interferometric configuration for entangled-multiphoton generation, our polarization-entangled four-photon GHZ state was robust and phase-stable [18].

The $|\Phi^+\rangle$ state of the four Bell states can be prepared by tuning the quarter-wave plate (QWP) and half-wave plate (HWP) positioned in each path. The input polarization-entangled states of Pair$_1$ and Pair$_2$ can be expressed as follows:

$$|\Phi^{in}\rangle = |\Phi^+\rangle_{Pair1} \otimes |\Phi^+\rangle_{Pair2}$$
$$= \frac{1}{\sqrt{2}}\left(|HH\rangle + |VV\rangle\right)_{Pair1} \otimes \frac{1}{\sqrt{2}}\left(|HH\rangle + |VV\rangle\right)_{Pair2} \quad (2)$$

When both idler photons are combined in and transmitted through the PBS, as shown in Fig. 1(b), the input state corresponding to Eq. (2) can be projected onto the four-photon polarization-entangled GHZ state with the same polarization states, as follows:

$$|^+GHZ^{(4)}\rangle = \frac{1}{\sqrt{2}}\left(|HHHH\rangle + |VVVV\rangle\right). \quad (3)$$

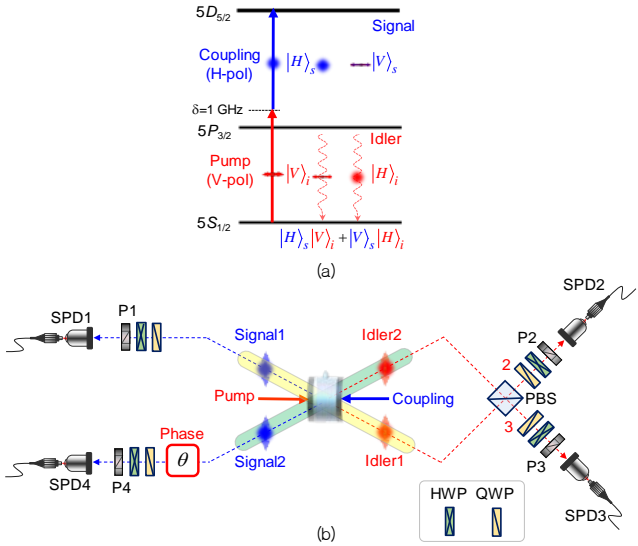

**Figure 1.** (a) Configuration for entangled photon-pair generation with $|H\rangle_s|V\rangle_i$ and $|V\rangle_s|H\rangle_i$ states via intrinsic polarization correlation in a three-level cascade-type atomic system. (b) Experimental configuration for the generation of a four-photon polarization-entangled Greenberger–Horne–Zeilinger (GHZ) state with phase $\theta$ and the use of two symmetrical spatial modes (Pair$_1$ and Pair$_2$) of inherently polarization-entangled photons from a warm $^{87}$Rb atomic vapor cell.

For practical purposes, entangled states can be prepared by operating on an arbitrary phase of genuine multipartite entanglement for a quantum network [29, 30], as in the case of entanglement distillation. Practically, we can consider a phase-GHZ state with phase $\theta$, as described in Eq. (4). In particular, it has been reported that the phase-GHZ state with a $\pi/4$ phase can be useful for magic-state distillation [28]. We can prepare the desired phase-GHZ state $|PGHZ^{(4)}(\theta)\rangle$ with phase $\theta$ by adjusting the local phase of one of the four photons. In this procedure, $|PGHZ^{(4)}(\theta)\rangle$ can be expressed as the unitary transformation of the $|GHZ^{(4)}\rangle$ state, with the zero-phase expressed in Eq. (3), as follows:

$$|PGHZ^{(4)}(\theta)\rangle = U(\theta)|GHZ^{(4)}\rangle$$
$$= \left[ I_1 \otimes I_2 \otimes I_3 \otimes \left(|H\rangle\langle H| + e^{i\theta}|V\rangle\langle V|\right)_4 \right]|GHZ^{(4)}\rangle$$
$$= \frac{1}{\sqrt{2}}\left(|HHHH\rangle + e^{i\theta}|VVVV\rangle\right),$$
(4)

where $U(\theta)$ denotes the unitary transformation operator with local phase $\theta$ in path4 in Fig. 1(b). In our experiment, we used an additional QWP in path4 as the phase-change device. We note here that the local phase $\theta$ can be tuned by adjusting the QWP angle. As shown in Fig. 1, the phase of the $|PGHZ^{(4)}(\theta)\rangle$ state is maintained because of the intrinsic polarization correlation achieved in a cascade-type atomic system with no interferometric configuration [20].

**Violation of the GBI of a phase-GHZ state**

The $S$-parameter in the GBI form of the four-partite GHZ state can be described as follows [1]:

$$S = \frac{1}{16}\sum_{s_1,\cdots s_4}\left|\sum_{k_1,\cdots k_4} f(s_1,\cdots s_4, k_1,\cdots k_4)\, E(k_1,\cdots k_4)\right|, \quad (5)$$

where $s_j = -1, 1$, $k_j = 1, 2$, the coefficient is given by $f(s_1\cdots k_4) = s_1^{k_1-1} s_2^{k_2-1} s_3^{k_3-1} s_4^{k_4-1}$, and the term of $E(k_1, k_2, k_3, k_4)$ denotes the correlation function (the average over many runs of the measurements). To obtain the quantum mechanical prediction of the correlation function, we considered polarization measurements for the four-photon GHZ states. Parameter $k_j$ denotes the measurement settings to be implemented such that the value of $k_j = 1$ corresponds to the linear polarization (D/A), where diagonal polarization $|D\rangle = 1/\sqrt{2}(|H\rangle+|V\rangle)$ and anti-diagonal polarization $|A\rangle = 1/\sqrt{2}(|H\rangle-|V\rangle)$, and $k_j = 2$ corresponds to the circular polarization (R/L), where right-circular polarization $|R\rangle = 1/\sqrt{2}(|H\rangle+i|V\rangle)$ and left-circular polarization $|L\rangle = 1/\sqrt{2}(|H\rangle-i|V\rangle)$. However, the classical prediction of the $S$-parameter in Eq. (5) by the local realistic model is given by $S_{LR} \leq 1$ and remains unchanged as a function of the phase $\theta$ of the GHZ state, indicated by the green line in Fig. 2.

In general, the maximal violations of the Bell inequalities are revealed by the interplay between the measurements and entangled states. Because the measurement settings are fixed in our working example, it is necessary to determine which entangled states can show the largest violation of the GBI. The values of $S$-parameter in Eq. (5) are functions of the phase $\theta$ of the $|PGHZ^{(4)}(\theta)\rangle$, which can be described as follows (see Supplementary Material for further details [31]):

$$S_{QM} = 2(|\cos\theta| + |\sin\theta|). \quad (6)$$

As indicated by the red curve in Fig. 2, the $S_{QM}$ values of the ideal four-partite GHZ state according to the phase of the $|PGHZ^{(4)}(\theta)\rangle$ periodically change, while the classical bound (green line in Fig. 2) remains unchanged. We find that $S_{QM}$ is enhanced by a factor of $\sqrt{2}$ at phases $\pi/4$ and $3\pi/4$, when compared with the ideal value of 2 at the zero phase and $\pi/2$.

To analyze the violations of the GBI of the $|PGHZ^{(4)}(\theta)\rangle$ state, we measured the fourfold coincidence counts of all 256 possible polarization combinations subjected to a projective polarization measurement basis. The experimental results in Fig. 2 at phases of 0, π/4, and π/2 were measured for a 60 min accumulation time with a 2.5 ns temporal window. For the first time, we observed the phase dependence of the $S_{QM}$ of the $|PGHZ^{(4)}(\theta)\rangle$ state by controlling the tilt angle of the QWP. In our experimental system, the polarization-entangled photon-pair coincident count rate was measured to be ~38,000 cps with a 2.5 ns coincidence window. The polarization-entangled four-photon GHZ state from the warm $^{87}$Rb atomic vapor cell was practically measured as 1.64(6) Hz under a weak pump power of 10 μW.

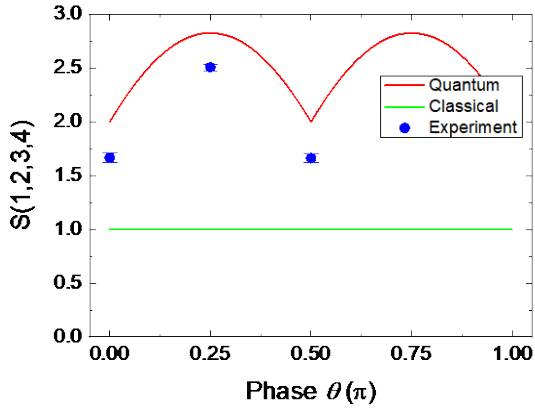

**Figure 2.** *S*-parameter in the GBI form as a function of the phase of the $|PGHZ^{(4)}(\theta)\rangle$ state from a warm atomic ensemble of $^{87}$Rb.

To obtain the value of $S_{QM}$, we measured 16 expectation values, as indicated in Eq. (5). Figure 3 shows the experimental results of the $E(k_1, k_2, k_3, k_4)$ measurements. In the $|PGHZ^{(4)}(\pi/4)\rangle$ state, the ideal value of $|E|$ is $1/\sqrt{2}$, as indicated by the height of the white bars shown for comparison with the experimental results (red bars). To measure the expectation value, the linear-polarization (D/A) and circular-polarization (R/L) measurements of the photons in the four paths can be obtained when we appropriately set the QWP and HWP corresponding to the GHZ state using two spatially multiplexed modes of the polarization-entangled photons from the warm atomic ensemble. Each polarization correlation of the four photons was measured within an accumulation time of 300 s. The experimental result of the $|PGHZ^{(4)}(\pi/4)\rangle$ state corresponding to Eq. (5) is estimated as $S_{QM} = 2.508 \pm 0.032$, which corresponds to the GBI violation of the $|PGHZ^{(4)}(\pi/4)\rangle$ state by approximately 47 standard deviations, as shown in Fig. 2. This result confirms the multipartite entanglement of the $|PGHZ^{(4)}(\pi/4)\rangle$ state. The phase-GHZ state is a genuine multipartite entanglement, which we subsequently confirmed by measuring the entanglement witness.

For the $|PGHZ^{(4)}(\theta)\rangle$ state at the zero phase and π/2, eight expectation values are dominant when measuring the *S*-parameter represented by Eq. (5), and the eight expectations are shown in Fig. 3. The $|PGHZ^{(4)}(0)\rangle$ state corresponds to the $|^{+}GHZ^{(4)}\rangle$ state of Eq. (3), and the eight dominant elements consist of an even number of linear polarization measurement settings, for example $E(1,1,1,1)$ or $E(1,1,2,2)$. For the π/2 phase case, the behavior is the opposite, such that the expectations consisting of an odd number of linear polarization settings are dominant. As shown in Fig. 2, the $S_{QM}$ value of the $|PGHZ^{(4)}(0)\rangle$ state is estimated to be $1.668 \pm 0.033$. We confirmed the genuine entanglement of the $|PGHZ^{(4)}(\theta)\rangle$ states from the violation of the GBI. In particular, the genuine entanglement of the $|PGHZ^{(4)}(\pi/4)\rangle$ state represents the strongest type of entanglement on the projective polarization measurement bases.

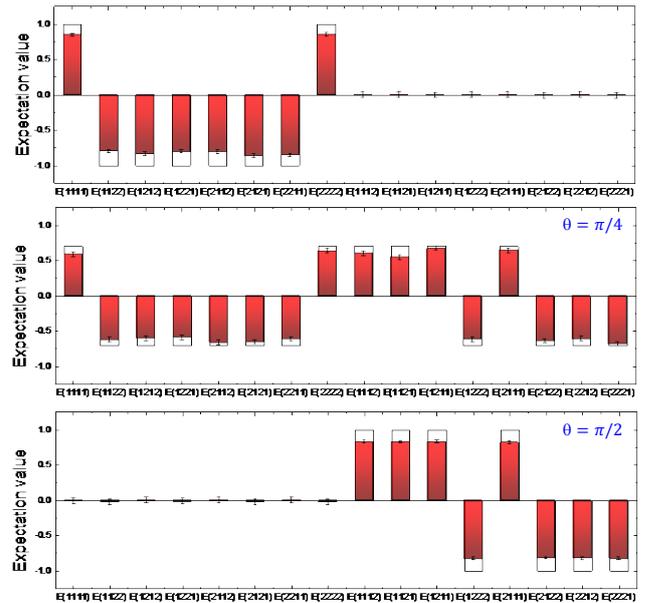

**Figure 3.** Experimental results observed in 16 specific measurements to obtain the expectation value from the four-photon polarization-entangled state $|PGHZ^{(4)}(\theta)\rangle$

at phases of 0, π/4, and π/2.

**QST of a phase-GHZ state**

The direct approach to characterize the GHZ state consists of performing full-state tomography, followed by an analysis of the reconstructed density matrix. To fully characterize the $\left|PGHZ^{(4)}(\theta)\right\rangle$ state with QST using projection measurements, the 256 components of the four-photon polarization-entangled GHZ state must be measured by adjusting the QWPs and HWPs in the paths (1, 2, 3, and 4 in Fig. 1(b)). We performed QST of the three $\left|PGHZ^{(4)}(\theta)\right\rangle$ states using the maximum likelihood estimation method, as shown in Fig. 4. The fidelity of the produced state is the extent to which the target state is prepared [32]. The fidelities of the $\left|PGHZ^{(4)}(\theta)\right\rangle$ states in Fig. 4 are estimated to range from 0.830 to 0.881 within an accumulation time of 300 s.

In our experimental system, the $\left|PGHZ^{(4)}(\theta)\right\rangle$ state can be robustly produced using an interferometric configuration owing to the intrinsic polarization correlation in the cascade-decay paths of the atomic transition. We can obtain the desired $\left|PGHZ^{(4)}(\theta)\right\rangle$ state, and the entanglement of the phase-GHZ states is independent of phase $\theta$. From the QST results, we can observe the transition of the reconstructed density matrix components for a phase range from zero to $\pi/2$.

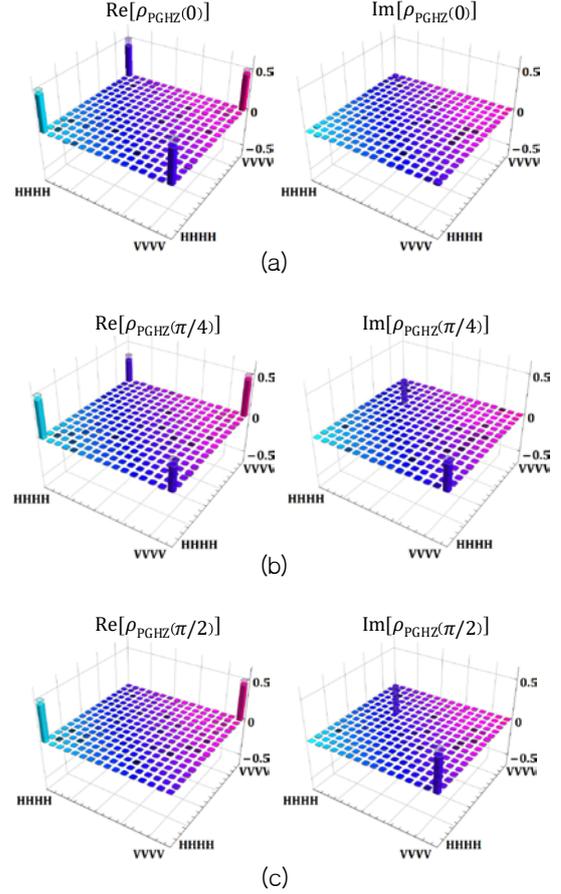

**Figure 4.** Real and imaginary components of the reconstructed density matrices $\rho_{PGHZ(\theta)}$ of the phase-GHZ states. (a) Fidelity value of 0.881 ± 0.02 in the $\left|PGHZ^{(4)}(0)\right\rangle$ state; (b) fidelity value of 0.842 ± 0.019 in the $\left|PGHZ^{(4)}(\frac{\pi}{4})\right\rangle$ state; and (c) fidelity value of 0.830 ± 0.018 of the $\left|PGHZ^{(4)}(\frac{\pi}{2})\right\rangle$ state.

**Entanglement witness of a phase-GHZ state**

Finally, we consider the witness to the genuine entanglement of the four-photon phase-GHZ state from the warm atomic ensemble. An observable $A$ is called an entanglement witness if $Tr(A\varrho_E) < 0$ for an entangled state $\varrho_E$ and $Tr(A\varrho_S) \geq 0$ for all separable states $\varrho_S$. The detection of negative values of the witness indicates that state $\varrho_E$ is entangled. QST and a generic witness for a pure state typically assume that the dimensions of the Hilbert space are known [22]. To verify whether the prepared state is a genuine multipartite entanglement state, a suitable entanglement witness can be used for entanglement detection and fidelity estimation.

Here, we employed the entanglement witness of a multipartite GHZ state, which has been studied theoretically and experimentally [22, 32, 33]. For the four-

photon phase-GHZ state, the witness can be expressed as

$$W_{PGHZ}(\theta) = \frac{I}{2} - |PGHZ^{(4)}(\theta)\rangle\langle PGHZ^{(4)}(\theta)| \quad (7)$$

where $I$ denotes the identity operator. The witness that is decomposed into tensor products of the local measurements is implemented experimentally (see Supplementary Material for further details [31]). Consequently, the expected values of $\hat{W}_{PGHZ}(\theta)$ for $\theta = 0$, $\pi/4$, and $\pi/2$ were estimated to be -0.375 ± 0.010, -0.404 ± 0.010, and -0.383 ± 0.008, respectively. The entanglement witnesses for all three situations were negative by more than 14 standard deviations, thus proving the occurrence of genuine four-partite entanglement.

## DISCUSSION

In summary, we experimentally demonstrated a genuine four-photon phase-GHZ state generated from a warm atomic ensemble of $^{87}$Rb atoms. In our experimental system, the phase-stable generation of the polarization-entangled phase-GHZ state could be achieved via the SFWM process of the cascade-type atomic transition owing to the intrinsic polarization correlation in the cascade-decay paths of the atomic transition. The fidelities of the phase-GHZ states obtained using a warm atomic vapor cell were estimated to be greater than 0.83 ± 0.02. To verify the genuine multipartite entanglement of our resource, we experimentally demonstrated the quantum violations of the Bell inequality and the high negative values of the entanglement witness. For this purpose, we employed the general Bell inequality (GBI) for $N$-qubit, derived by Żukowski and Brukner [1], which recovers all possible local realistic descriptions of $N$-partite systems; for example, one can reproduce the MABK inequality from the one considered here. It has been shown that a large class of generalized GHZ states can violate the general Bell inequality where the MABK inequality fails [34, 35]. Our Bell test shows quantum violations for the phase-GHZ state with phase π/4 by 47 standard deviations and also shows high negative values for the four-photon entanglement witness, which in turn confirms that the phase-GHZ state is a genuine multipartite entanglement state. Our results can contribute to realizing a novel resource for assisting magic distillation processes, multipartner secret sharing, controlled cryptography, and quantum networks based on atom–photon interactions.

## METHODS
### Experimental details
In our experiment, a polarization-entangled four-photon GHZ state was realized by combining photons from two different spatial modes of the polarization-entangled photon-pair source resulting from the $5S_{1/2}$–$5P_{3/2}$–$5D_{5/2}$ transition of $^{87}$Rb atoms, as shown in Fig. 1. Under the two-photon resonant condition interacting with the cascade-type atomic system with orthogonally linearly polarized pump and coupling lasers, a strongly correlated signal and idler photon pair are emitted via the SFWM process owing to the collective two-photon coherence in the Doppler-broadened atomic ensemble of $^{87}$Rb. To maximize the two-photon coherence between the $5S_{1/2}(F = 2)$ and $5D_{5/2}(F'' = 4)$ states and suppress the single-photon resonances of the $5S_{1/2}(F = 2)$–$5P_{3/2}$ transition for the pump laser and the $5P_{3/2}$–$5D_{5/2}$ transition for the coupling laser, the detuning frequency (δ) of the pump and coupling lasers is set at ~1 GHz, beyond the Doppler-broadening of the warm $^{87}$Rb atoms. The vertically polarized (V-pol) pump and horizontally polarized (H-pol) coupling lasers induce two-photon coherence in the cascade-type atomic system, as shown in Fig. 1(a).

## Supplementary material

See the Supplementary Material for detailed descriptions of the characteristics of the single general Bell inequality for the four parties and the entanglement witness of the four-photon GHZ state.

## Data availability

Data are available from the corresponding authors upon reasonable request.

## Acknowledgment

This study was supported by the National Research Foundation of Korea (NRF) (Grant Nos. NRF-2021R1A2B5B03002377 and NRF-2020M3E4A1080030) and the Ministry of Science and ICT (MSIT), Korea, under the Information Technology Research Center (ITRC) support program (IITP-2023-2020-0-01606) and (Grant No. 2022-0-01029) supervised by the Institute of Information & Communications Technology Planning & Evaluation (IITP). J.R. acknowledges the National Research Foundation of Korea (NRF) (Grant Nos. NRF-2020M3E4A1079792 and NRF-2022M3K2A1083890).

## Author contributions

H. S. M. planned and supervised the research; J. P. and J. R. carried out the experiment and the theoretical calculations; J. P. and H. S. M. analyzed the data; J. P., J. R., H. K., and H. S. M. contributed the discussion of the results; H. K., J. R., H. K., and H. S. M. wrote the manuscript with input from all authors.

**Competing interests**

The authors declare no competing financial or non-financial interests.

# Supplementary Information

Jiho Park[1], Junghee Ryu[2], Heonoh Kim[1], and Han Seb Moon[1]

[1] *Department of Physics, Pusan National University, Geumjeong-Gu, Busan 46241, Korea*
[2] *Division of National Supercomputing, Korea Institute of Science and Technology Information, Daejeon 34141, Republic of Korea*

## 1. Single general Bell inequality for four parties

Consider four parties and allow each to select independently between two different observables, where each observable has two outcomes. A local realistic description implies that the outcomes of the observables are predetermined before the measurements are implemented. We take as our working example the single general Bell inequality introduced in [1], which reads as follows:

$$S = \sum_{s_1,\cdots s_4 = 1,-1} \left| \sum_{k_1,\cdots k_4 = 1,2} s_1^{k_1-1} s_2^{k_2-1} s_3^{k_3-1} s_4^{k_4-1} E(k_1,\cdots k_4) \right| \leq 2^4, \tag{S1}$$

where $E(k_1, k_2, k_3, k_4)$ denote the averages of many experimental runs. Note that in the main text, the *S*-parameter in (S1) is normalized by its classical upper bound, that is, 16, such that the maximum of the local realistic description of the *S*-parameter is given by one. A four-qubit Greenberger–Horne–Zeilinger (GHZ) state with a phase in the form

$$\left| PGHZ^{(4)}(\theta) \right\rangle = \frac{1}{\sqrt{2}} \left( \left| HHHH \right\rangle + e^{i\theta} \left| VVVV \right\rangle \right) \tag{S2}$$

shows the quantum violation of the *S*-parameter. To obtain the values of $E(k_1, k_2, k_3, k_4)$, we implemented polarization measurements such that $k_j = 1$ corresponds to the linear polarization (D/A) measurement and $k_j = 2$ to the circular polarization (R/L) measurement. Consequently, the *S*-parameter can be expressed as

$$S_{QM} = 2 \left( |\cos\theta| + |\sin\theta| \right), \tag{S3}$$

where we use the following rules: if the number of $k_j = 1$ instances is even, then $E(k_1, k_2, k_3, k_4) = \pm \cos(\theta)$ and if the $k_j = 1$ instances are odd, then $E(k_1, k_2, k_3, k_4) = \pm \sin(\theta)$.

## 2. Entanglement witness of four-photon GHZ state

We experimentally implemented an entanglement witness for the four-photon polarization-entangled phase-GHZ state, which aided in confirming that the state generated in our laboratory is genuinely multipartite entanglement. To this end, we used the following entanglement witness [2]:

$$W_{PGHZ}(\theta) = \frac{I}{2} - \left| PGHZ^{(4)}(\theta) \right\rangle \left\langle PGHZ^{(4)}(\theta) \right|, \tag{S4}$$

where *I* denotes the identity operator. Note that to implement the witness experimentally, the witness operator may need to be decomposed into the tensor product of the local measurement operators. For $\theta = 0$, the witness reads:

$$W_{PGHZ}(0) = \frac{I}{2} - |PGHZ^{(4)}(0)\rangle\langle PGHZ^{(4)}(0)|$$
$$= \frac{I}{2} - \frac{1}{2}\left(|H\rangle\langle H|^{\otimes 4} + |V\rangle\langle V|^{\otimes 4} + |H\rangle\langle V|^{\otimes 4} + |V\rangle\langle H|^{\otimes 4}\right) \quad \text{(S4)}$$

The last two terms can be represented by the following correlation functions:

$$\frac{1}{8}\begin{pmatrix} E(1, 1, 1, 1) - E(2, 2, 1, 1) - E(1, 2, 2, 1) - E(2, 1, 2, 1) \\ -E(1, 2, 1, 2) - E(2, 1, 1, 2) - E(1, 1, 2, 2) + E(2, 2, 2, 2) \end{pmatrix} \quad \text{(S5)}$$

For $\theta = \pi/4$, similarly, we have

$$W_{PGHZ}\left(\frac{\pi}{4}\right) = \frac{I}{2} - \left|PGHZ^{(4)}\left(\frac{\pi}{4}\right)\right\rangle\left\langle PGHZ^{(4)}\left(\frac{\pi}{4}\right)\right|$$
$$= \frac{I}{2} - \frac{1}{2}\left(|H\rangle\langle H|^{\otimes 4} + |V\rangle\langle V|^{\otimes 4} + \left(\frac{1-i}{\sqrt{2}}\right)|H\rangle\langle V|^{\otimes 4} + \left(\frac{1+i}{\sqrt{2}}\right)|V\rangle\langle H|^{\otimes 4}\right), \quad \text{(S6)}$$

and moreover, the last two terms read

$$\frac{1}{8\sqrt{2}}\begin{pmatrix} E(1, 1, 1, 1) - E(2, 2, 1, 1) - E(1, 2, 2, 1) - E(2, 1, 2, 1) \\ -E(1, 2, 1, 2) - E(2, 1, 1, 2) - E(1, 1, 2, 2) + E(2, 2, 2, 2) \\ +E(1, 2, 1, 1) + E(2, 1, 1, 1) + E(1, 1, 2, 1) - E(2, 2, 2, 1) \\ +E(1, 1, 1, 2) - E(2, 2, 1, 2) + E(1, 2, 2, 2) + E(2, 1, 2, 2) \end{pmatrix}. \quad \text{(S7)}$$

For $\theta = \pi/2$,

$$W_{PGHZ}\left(\frac{\pi}{2}\right) = \frac{I}{2} - \left|PGHZ^{(4)}\left(\frac{\pi}{2}\right)\right\rangle\left\langle PGHZ^{(4)}\left(\frac{\pi}{2}\right)\right|$$
$$= \frac{I}{2} - \frac{1}{2}\left(|H\rangle\langle H|^{\otimes 4} + |V\rangle\langle V|^{\otimes 4} - i|H\rangle\langle V|^{\otimes 4} + i|V\rangle\langle H|^{\otimes 4}\right) \quad \text{(S8)}$$

and

$$\frac{1}{8}\begin{pmatrix} E(1, 2, 1, 1) + E(2, 1, 1, 1) + E(1, 1, 2, 1) - E(2, 2, 2, 1) \\ +E(1, 1, 1, 2) - E(2, 2, 1, 2) + E(1, 2, 2, 2) + E(2, 1, 2, 2) \end{pmatrix}. \quad \text{(S9)}$$

The experimental data obtained based on the above relations are summarized in Table 1.

Table 1. Experimental results of the entanglement witness

| $\theta$ | $W_{PGHZ}(\theta)$ |
|---|---|
| 0 | $-0.375 \pm 0.010$ |
| $\pi/4$ | $-0.404 \pm 0.010$ |
| $\pi/2$ | $-0.383 \pm 0.008$ |

For the three different phases of $\theta$, negative values for the entanglement witness were detected, involving more than 14 standard deviations. This result implies that the generated state is a genuine four-partite entanglement state.